\documentclass[11pt]{article}
\usepackage[top=1in, bottom=1in, left=1in, right=1in]{geometry}
\usepackage{amsmath}
\usepackage{mathrsfs}
\usepackage{epsfig}
\usepackage{graphicx}
\usepackage{authblk}
\usepackage{cite}
\usepackage{url}
\usepackage{hyperref}

\usepackage{color}

\begin{document}

\title{Time-dependent influence metric for cascade dynamics on networks}

\author[1]{James~P.~Gleeson}
\author[1]{Ailbhe Cassidy} 
\author[2]{Daniel Giles} 
\author[3]{Ali Faqeeh}
\affil[1]{MACSI, Department of Mathematics \& Statistics, University of Limerick, Ireland}
\affil[2]{Department of Computer Science, University College London, UK}
\affil[3]{Department of Computer Science, Aalto University, Finland}

\date{ }

\maketitle
\begin{abstract}
    An algorithm for efficiently calculating the expected size of single-seed cascade dynamics on networks is proposed and tested. The expected \textcolor{black}{cascade} size is a time-dependent quantity and so enables the identification of nodes who are the most influential early or late in the spreading process. The measure is accurate for both critical and subcritical dynamic regimes and so generalises the nonbacktracking centrality that was previously shown to successfully identify the most influential single spreaders in a model of critical epidemics on networks.
\end{abstract}

\section{Introduction}\label{sec:Intro}
The challenge of identifying influential nodes in a network has been intensively studied. Many measures have been proposed to predict the importance or centrality of nodes, the majority of such methods being based (solely) on the connectivity or topology of the network. In the era of big data, however, it is increasingly possible to observe interactions at a node-to-node level, e.g., the likelihood that node $j$ will retweet a message they received from node $i$. Moreover, the dynamics of retweeting (and of related node-to-node interactions) are strongly affected by the temporal dynamics of human behaviour.

In this paper we consider how influential a node is in terms of its ability to propagate information through the network in a timely fashion. To see why such a measure can differ from classical centrality measures, consider a node $i$ who has many Twitter followers (i.e., other nodes who receive the messages tweeted by node $i$).
Any connectivity-based measure will deem such a node as influential, but this influence depends on the followers of node $i$ being willing to retweet its messages, and their followers in turn also transmitting the message further. If for some reason the followers of node $i$ all decide not to retweet messages from them, or if their response time is relatively slow, then some other node with a lower number of (more active) followers may in fact be superior in terms of broadcasting reach into the network. In this paper we introduce a dynamical measure for the influence of a node in a network where transmission of a message or an infection from node $i$ to node $j$ happens with a time-dependent rate $\psi_{i\to j}(t)$ that may be different for each pair of nodes, and may also depend on the time $t$ elapsed since node $i$ became infected (received the tweet).
In terms of Twitter, for example, $\psi_{i\to j}(t)$ could be calculated empirically for each friend-follower relationship $(i,j)$ by observing the fraction of tweets from node $i$ that are retweeted by node $j$ in a certain timeframe.

We develop a measure for the influence of each node by calculating the expected size of the ``cascade tree'' that is seeded by the infection of the node, under  \textcolor{black}{  locally tree-like  and infinite-size assumptions on the network structure and under branching-process assumptions on the dynamics. The tree size measure} is the total number of nodes that retweet a message originating at node $i$ (plus one, for the seed node itself). The measure is time-dependent, so that we can identify those nodes who quickly generate cascades (perhaps due to having particularly active followers) and compare the influence of nodes at both early and late times in the cascades. Although our derivation of the measure is based on a tree-like assumption for the network, we show that our influence metric works well even on real-world networks. We focus on the scalability of the algorithm with network size, and propose a numerically efficient approach that makes calculation of the tree-size influence measure feasible even in very large networks.

This work was motivated and inspired by the paper \cite{Radicchi16a} by Radicchi and Castellano, who show that the nonbacktracking centrality identifies the most influential single spreaders in a susceptible-infected-recovered (SIR) model of epidemic spreading at criticality. Our work extends \cite{Radicchi16a} in several directions: we examine the general class of subcritical spreading processes, taking the spreading at criticality as a limiting case, and reproducing the results of \cite{Radicchi16a} in that special limit case. Moreover, we consider time-dependent influence, with the relative importance of nodes potentially being different at short times than in the long-time or static limit that is considered in \cite{Radicchi16a} and most other papers.

Other works that have examined single-node influence metrics from related viewpoints include \cite{larremore2012statistical} and \cite{friedkin16}. Both of these papers consider influence metrics that are calculated form the (weighted) adjacency matrix, rather than using the weighted nonbacktracking matrix as we do here, and so give different results on undirected networks and on directed networks with reciprocal edges (see Fig.~\ref{figrealworld}). 
\textcolor{black}{
Reference \cite{min2018identifying} presents an approach that enables the determination of node influence for SIR dynamics in the long-time limit, using a message-passing solution for bond percolation. The focus of \cite{min2018identifying} is on global epidemics, i.e., supercritical dynamics where infection spreads to a substantial fraction of the whole network (the giant connected component) and their work also reproduces the results of \cite{Radicchi16a} in the limiting case of critical spreading. We focus here on subcritical dynamics (corresponding to small outbreaks in the SIR disease context) and emphasize the time-dependency of the resulting influence metric.
}

The approach we study here is restricted to individual spreaders, i.e., to the question of what individual node can seed the largest cascade or outbreak. Substantial effort has  been devoted to the separate optimisation problems of determining the set of optimal multiple spreaders, where the question is to identify a subset (of given size) of nodes that, when seeded simultaneously, create the largest cascade, see, for example \cite{Kempe03,braunstein2016network,morone2015influence,morone2016collective,sen2017efficient}. Knowing the most influential single spreaders does not solve the multiple-spreader problems because of overlaps between the cascades of the individual seeds when they start simultaneously.
While the individual spreader problem studied here is in some sense easier that the multiple-seed question, we believe that our contribution will prove useful in further elucidating spreading mechanisms and particularly the role of time-dependence in influence rankings.

The remainder of this paper is structured as follows. In Sec.~\ref{sec:metric} we define our metric, which we call the ``tree-size influence''. In Sec.~\ref{sec:ICM} we apply it to the discrete-time Independent Cascade Model at various times and compare with some other influence metrics, including the non-backtracking centrality used in \cite{Radicchi16a}. A continuous-time model for the propagation of memes on directed networks like Twitter is used in Sec.~\ref{sec:memeprop} to illustrate the time-dependence of the influence rankings. In Sec.~\ref{sec:realworld} we apply the tree-size influence metric to cascades simulated on real-world networks and compare with further centrality measures. Conclusions are drawn in Sec.~\ref{sec:conclusions} and four short appendices give further mathematical details.




\section{Tree-size influence measure} \label{sec:metric}
We consider unidirectional (or ``monotone'') binary-state dynamics, where at each instant a node is in either the ``susceptible'' state or the ``infected'' state. Once infected, a node cannot return to the susceptible state. For example, if we are tracking the propagation of a specific tweet on Twitter, we label a node as infected if it has tweeted or retweeted the message, and susceptible if it has not done so.
We define $\psi_{i\to j}(\tau)\, d\tau$ as the probability that a transmission along directed edge\footnote{We adopt the convention that edges are directed from a node to its followers, i.e., in the direction of the flow of information.} $i\to j$ occurs at a time in the interval $[\tau, \tau+d\tau]$ after the moment of infection of node $i$ \cite{Karrer10}. In some models of Twitter dynamics, for example, the probability of node $j$ retweeting a message it has received depends inversely on the in-degree of node $j$ (i.e., the higher the number of Twitter users $j$ follows, the less likely he is to retweet a particular message), due to the competition between incoming tweets for the limited attention of user $j$ \cite{Lerman16,Weng12}.

Our tree-size influence metric for node $i$ is the expected size of the cascade initiated by the infection of node $i$, as measured at a time $t$ after the initial infection. The cascade is modelled as a tree, where the infection of each node leads to the possible infection of its (out-)neighbours. In the Twitter example, the expected tree size for node $i$ is the expected number of retweets of a message broadcast by node $i$ (plus one, to count the original tweeting by node $i$ itself).

Suppose node $i$ is infected. Define $m_{i\to j}(t)$ to be the expected (mean) size, at a time $t$ after the infection of node $i$, of the tree created by transmission along the edge from node $i$ to node $j$.
\textcolor{black}{
We assume (as is common in branching-process approximations) that the subtrees created by the infection of two neighbours $j$ and $k$ of node $i$ are independent of each other.} Noting that the probability that infection is transmitted from $i$ to $j$ by time $t$ is given by the integral of $\psi_{i \to j}(\tau)$ from 0 to $t$, we can then write an integral equation for $m_{i \to j}(t)$:
\begin{equation}
m_{i \to j}(t) = \int_0^t \psi_{i \to j}(\tau) \left[ 1 + \sum_{k\in {\mathcal{N}_j}\setminus i} m_{ j \to k}(t-\tau)\right] \, d\tau \label{1},
\end{equation}
where ${\mathcal{N}_j}$ is the set of neighbours (out-neighbours if the network is directed) of node $j$.
The first term in the square brackets counts the contribution of 1 to the cascade size by node $j$ becoming infected, and the second term adds the sizes of the subcascades initiated by subsequent infections starting from node $j$. If $j$ is infected at time $\tau$ then the subtree has age $t-\tau$ when time is $t$. Then the expected size of a tree started by the infection of node $i$ depends on the the time $t$ elapsed since $i$ was infected as
\begin{equation}
s_i(t) = 1 + \sum_{j\in {\mathcal{N}_i}} m_{i \to j}(t). \label{2}
\end{equation}
\textcolor{black}{
Note that the branching-process assumption of independence of ``child'' subtrees can provide a good approximation for subcritical dynamics on finite-sized and tree-like networks, where the cascade does not grow to a substantial fraction of the network and the $m_{i \to j}$ terms in Eq.~(\ref{1}) remain finite. Our analysis is therefore limited to subcritical and critical dynamics.
}

\subsection{Long-time limit}
Let $p_{i \to j}$ be the probability that the infection of node $i$ eventually leads to the infection of node $j$ (cf.~the \emph{vulnerability} defined in \cite{GleesonDurrett17}):
\begin{equation}
p_{i\to j} = \int_0^\infty \psi_{i \to j}(\tau) \, d\tau \label{vpeqn}.
\end{equation}
Taking the limit $t\to\infty$ of Eq.~(\ref{1}) yields a linear system of algebraic equations \textcolor{black}{for the quantities $m_{i\to j}(\infty)$}:
\begin{equation}
m_{i\to j}(\infty) = p_{i \to j}\left[ 1 + \sum_{k\in {\mathcal{N}_j}\setminus i} m_{ j \to k}(\infty)\right], \label{3}
\end{equation}
with Eq.~(\ref{2}) similarly limiting to
\begin{equation}
s_i(\infty) = 1 + \sum_{j\in {\mathcal{N}_i}} m_{i \to j}(\infty). \label{4}
\end{equation}
Note that Eq.~(\ref{3}) can be written as
\begin{equation}
m_{i\to j}(\infty) = p_{i \to j}  +  p_{i \to j} \sum_{\text{edges} j\to k} B_{i\to j, j\to k} m_{ j \to k}(\infty), \label{3b}
\end{equation}
where $B_{i\to j, j\to k}$ is the nonbacktracking (Hashimoto) matrix \cite{Karrer14} of the network.

The vector $\mathbf{m}$ of values of $m_{i \to j}(\infty)$ is the solution of the matrix equation
\begin{equation}
\mathbf{m} = \mathbf{p}+ \mathbf{P} \mathbf{B} \mathbf{m},
\end{equation}
where $\mathbf{p}$ is the vector of $p_{i \to j}$ values, $\mathbf{B}$ is the nonbacktracking matrix, and $\mathbf{P}$ is a  matrix with the vector $\mathbf{p}$ on its diagonal.
This can be rewritten as the linear system
\begin{equation}
\left(\mathbf{I}_E - \mathbf{P}\mathbf{B}\right) \mathbf{m} = \mathbf{p}, \label{matrixeq}
\end{equation}
where $\mathbf{I}_E$ is the identity matrix of dimension $E$ ($E$ is the number of directed edges: each undirected edge (e.g., $i$--$j$) of the original network is replaced by two reciprocal directed edges (e.g., $i\to j$ and $j\to i$), so for an undirected network with $e$ edges, $E=2e$).
In Appendix~\ref{appA} we consider numerical methods for efficient solution of  Eq.~(\ref{matrixeq}).

\subsection{Time-dependent case}
In simple discrete-time models of contagion, such as the Independent Cascade Model \cite{Kempe03} , all nodes are synchronously updated in every time step \textcolor{black}{and infected nodes recover after one time step (see Sec.~\ref{sec:ICM}). The transmission rate is therefore}
\begin{equation}
\psi_{i\to j }(t) = p_{i \to j} \, \delta(t-1),
\end{equation}
where $\delta$ is the Dirac delta function.
\textcolor{black}{
In this case the quantity
$p_{i \to j}$ defined in Eq.~(\ref{vpeqn}) equals the probability of a successful infection transmission from node $i$ to node $j$ in the timestep following the infection of node $i$ (since node $i$ recovers after one time step, the transmission along the $i\to j$ network link either happens in the time step following infection of node $i$, or not at all).} 

For such models, Eq.~(\ref{1}) becomes
\begin{equation}
m_{i\to j}(t) = p_{i \to j} \left[ 1 + \sum_{k\in {\mathcal{N}_j}\setminus i} m_{ j \to k}(t-1)\right] \quad\text{ for } t=1,2,\ldots, \label{it}
\end{equation}
with $m_{i\to j}(0)=0$ for all edges. This equation can be iterated to determine $m_{i\to j}(t)$ after $t$ discrete time steps, with  the expected tree sizes then given by Eq.~(\ref{2}).

More generally, Eq.~(\ref{1}) can be Laplace transformed to
\begin{equation}
\widehat{m}_{i\to j}(s) = \frac{1}{s} {\widehat{\psi}}_{i\to j}(s) + {\widehat{\psi}}_{i\to j}(s) \sum_{k\in {\mathcal{N}_j}\setminus i} \widehat{m}_{j\to k}(s), \label{LT}
\end{equation}
where hats denote Laplace transforms, e.g., $\widehat{m}_{i \to j}(s) = \int_0^\infty e^{-s t} m_{i\to j}(t) dt$. For any fixed value of $s$, Eq.~(\ref{LT}) can be written as a matrix equation similar to Eq.~(\ref{matrixeq}) and numerically solved using the methods described in Appendix~\ref{appA}. The efficient Talbot method for numerical inversion of Laplace transforms (see \cite{Abate06} and the summary in Appendix~\ref{appD}) can then be applied to obtain $m_{i \to j}(t)$ at any $t$ value by solving the system (\ref{LT}) at a finite (and relatively small) number of values of $s$; we give examples in Sec.~\ref{sec:memeprop} below.

\section{Independent Cascade Model}\label{sec:ICM}
We begin with a simple, discrete-time infection model.
 Specifically, we focus on the  Independent Cascade Model \cite{Kempe03} (as also used in \cite{Radicchi16b}), which is essentially a Reed-Frost disease-spread model on an undirected network, i.e., a discrete-time version of the continuous-time SIR epidemic model. The cascade starts from one infected node. In each discrete time step, all infected nodes contact each of their uninfected neighbours, and succeed in transmitting the infection to each neighbour (independently) with a common probability $p$. Having made one attempt to pass the infection to each neighbour, the node is then labelled as recovered and cannot become infected again.

The expected tree sizes at time $t$ are calculated from Eq.~(2), where the edge-based quantities $m_{i \to j}(t)$ are found by iterating Eq.~(\ref{it}), using the values $p_{i \to j}\equiv p$ for the Independent Cascade Model dynamics.

We start by considering the first step of the cascade process, and set $t=1$ in Eq.~(\ref{it})  to obtain $m_{i\to j}(1)=p$ for all edges $i\to j$. The corresponding node influence values are therefore given by Eq.~(\ref{2}) as
\begin{equation}
s_i(1)= 1 + p \,\, \text{deg}(i), \label{eq12}
\end{equation}
where $\text{deg}(i)$ is the degree of node $i$. It immediately follows that the ranking of nodes by tree size at $t=1$ is exactly the same as ranking the nodes by their degree.

Next, we take the limit $t\to \infty$ in Eqs.~(\ref{2}) and (\ref{it}). The vector $\mathbf{m}$ of values $m_{i \to j}(\infty)$ then is the solution of the matrix equation
\begin{equation}
\left( \mathbf{I}_E - p \mathbf{B}\right) \mathbf{m} = p \mathbf{1}, \label{prehvect}
\end{equation}
with solution
\begin{equation}
\mathbf{m} = p \left(\mathbf{I}_E-p \mathbf{B}\right)^{-1} \mathbf{1},\label{hvect}
\end{equation}
where $\mathbf{1}$ is the vector with all components equal to 1. The inverse matrix used in Eq.~(\ref{hvect}) is singular when the value of $p$ equals the critical value $p_c = 1/\lambda_\text{max}$, where $\lambda_\text{max}$ is the largest eigenvalue of the nonbacktracking matrix $\mathbf{B}$. We note that Eq.~(\ref{prehvect}) is precisely the same equation found from the analysis in \cite{Karrer14}  for the expected size of the (subcritical) percolation cluster reachable along the edge from $i$ to $j$ (our $m_{i\to j}(\infty)$ is their $H_{i \leftarrow j}^\prime(1)$). In the limit as $p$ approaches $p_c$ from below, it can be shown
that the vector $\mathbf{m}$ given by Eq.~(\ref{hvect}) is proportional to the leading eigenvector of the nonbacktracking matrix. We therefore conclude that in the limit of $t\to \infty$, and if $p$ is near its critical value $p_c$, that the tree size influence ranking will coincide with the ranking given by the nonbacktracking centrality defined in \cite{Martin14}. The nonbacktracking centrality measure was  shown to outperform other commonly-used metrics in locating influential seeds for the SIR disease-spread process at criticality \cite{Radicchi16a}.

\begin{figure}
\centering
\epsfig{figure=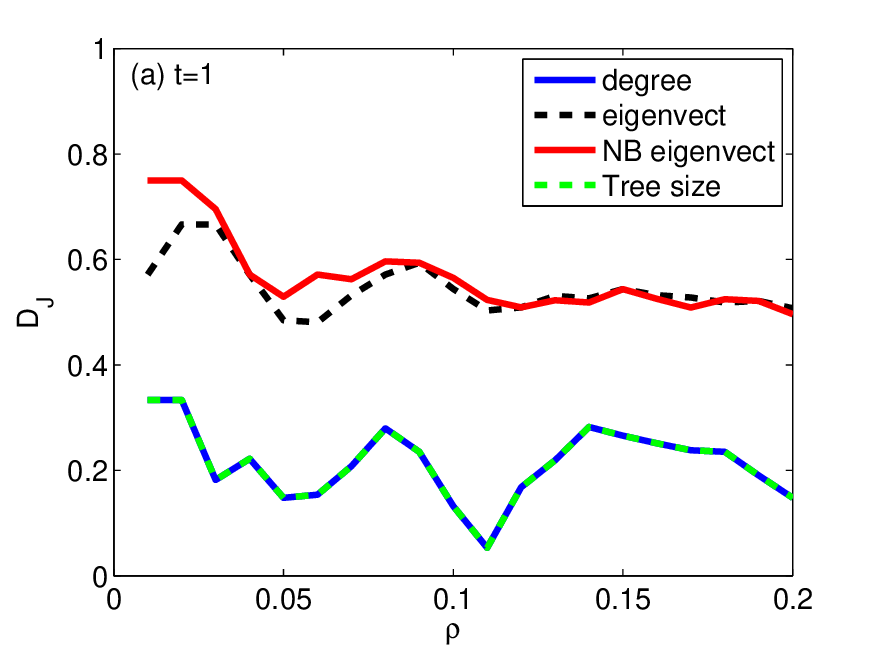,width=8.1 cm} 
\epsfig{figure=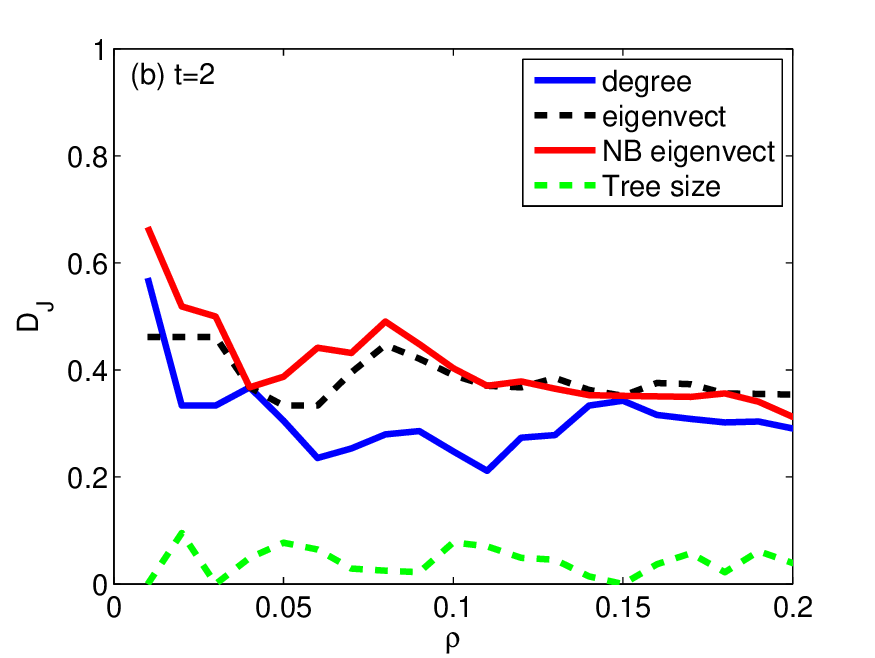,width=8.1 cm}
\epsfig{figure=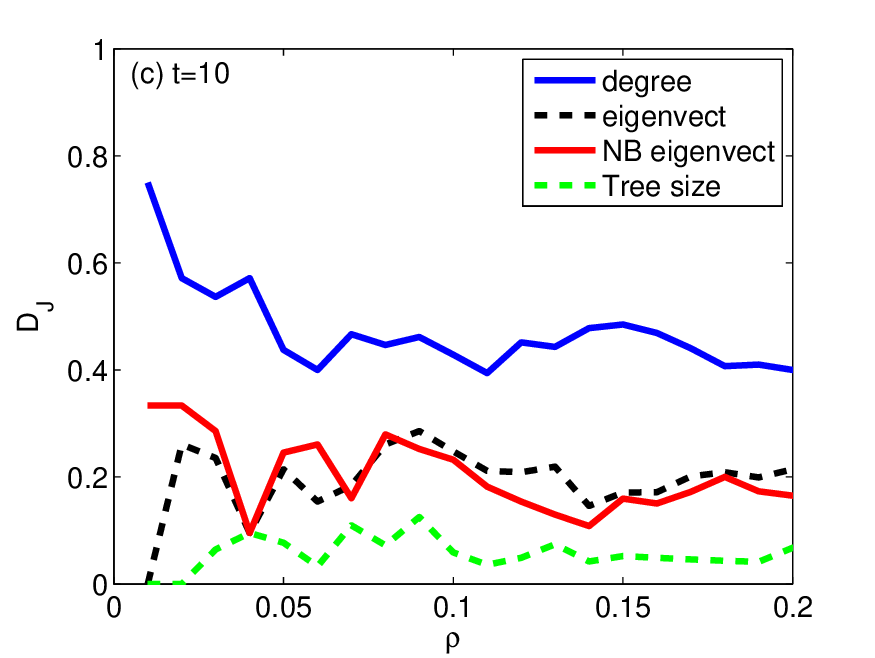,width=8.1 cm}
\epsfig{figure=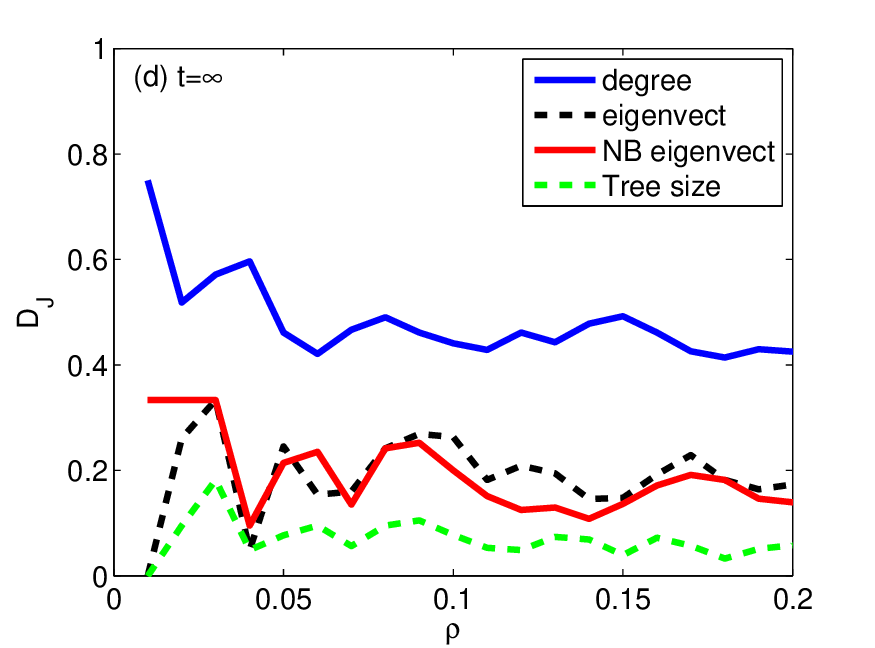,width=8.1 cm}
\caption{Jaccard distance $D_J$ between the set of nodes corresponding to the top faction $\rho$ of nodes  as given by the various influence metrics, and  as determined from numerical simulations of the Independent Cascade Model with $p=0.8 p_c$ up to times $t$, with (a) $t=1$, (b) $t=2$, (c) $t=10$, (d) $t=\infty$. Lower values of $D_J$ represent the more accurate metrics; note the tree size influence (as determined by Eqs.~(\ref{2}) and (\ref{it})) is the best in all cases (it coincides with the degree centrality at $t=1$).
\textcolor{black}{It is intersting to note that the tree size metric is more accurate (i.e., has lower $D_J$ values) for later times than for $t=1$. This is because at $t=1$ there is only a one-step infection process, meaning that all nodes with the same degree will have exactly the same prediction for their cascade size. In the numerical simulations, there will be small fluctuations in the expected sizes of cascades, leading to a different ranking of nodes, and hence to a larger Jaccard distance between predicted and actual ranks of nodes with the same degree.}
}\label{figsubcrit}
\end{figure}

To test the predictions of our theory,
we run a large number of simulations of the Independent Cascade Model on a single undirected Erd\H{o}s-R\'{e}nyi graph with $N=10^3$ nodes and mean degree $z=4$. Figure~\ref{figsubcrit} shows results for the subcritical case, where the transmission parameter $p$ is equal to $0.8 p_c$. Each node of the network is used as the seed for $10^4$ realizations of the process, and the average size of the resulting cascade (averaged over the $10^4$ realizations) is recorded at times $t=1$, $t=2$, $t=10$, and when the cascade has terminated ($t=\infty$). The nodes can therefore be ranked by the average size of the cascades they seed, and these rankings change as the observation time $t$ is increased. We compare the rankings obtained from the numerical simulations with the rankings predicted by four commonly-used influence metrics:
\begin{enumerate}
\item Degree (blue solid line): nodes are ranked according to their degree.
\item Eigenvector centrality (black dashed line): nodes are ranked according to the corresponding values in the leading eigenvector of the adjacency matrix of the network.
\item Nonbacktracking centrality (red solid line): nodes are ranked according to the corresponding values in the leading eigenvector of the nonbacktracking matrix \cite{Martin14,Radicchi16a}.
\item Tree size influence (green dashed line): nodes are ranked according to the values $s_i(T)$ calculated from Eqs.~(\ref{2}) and (\ref{it}).
\end{enumerate}
Note that measures 1 through 3 are structural metrics and so do not change with the evolution of the cascade process, in contrast to the tree size influence metric which (as shown in Sec.~\ref{sec:metric}) gives rankings that change in time.

\textcolor{black}{As in  \cite{Radicchi16a},} the accuracy of the rankings given by the four metrics are compared by calculating the Jaccard distance $D_J$  between the top fraction $\rho$ of nodes as ranked by the  (i) cascade size seeded by that node as predicted by the metric  and (ii) the ground truth average cascade size seeded by the node as calculated in the numerical simulations. 
The Jaccard distance measures the dissimilarity between two sample sets $A$ and $B$ as the complement of the Jaccard coefficient:
\begin{equation}
    D_J = 1 - \frac{\left|A\cap B\right|}{\left|A\cup B\right|},
\end{equation}
where $\left|A\right|$ denotes the number of elements in set $A$. Following \cite{Radicchi16a}, we take the top fraction $\rho$ of nodes as ranked by an influence metric to constitute set $A$, and (for the same value of $\rho$) we calculate the Jaccard distance to the top fraction $\rho$ of nodes as determined by numerical simulations of the spreading process . If the two sets $A$ and $B$ are equal then the influence metric is correctly predicting every node in the top fraction $\rho$ of spreaders, and the corresponding value of $D_J$ is zero. We consider a range of values of $\rho$ to determine the robustness of the measure: the best influence measure should have low values of $D_J$ across the range of $\rho$ values, meaning that the metric captures the ranking of nodes by influence not just for the most influential nodes but also ranks accurately those nodes that are, for example, in the top $20\%$ of influencers.

Figure~\ref{figsubcrit} shows the Jaccard distances $D_J$ for each of the four metrics at times $t=1$, $t=2$, $t=10$ and $t=\infty$ \textcolor{black}{with the transmission parameter $p$ fixed at the value $p=0.8 p_c$.} At $t=1$ (Fig.~\ref{figsubcrit}(a)), the tree size influence metric gives exactly the same ranking of nodes as degree centrality (note the green dashed line is exactly overlaid on the blue solid line), as predicted by Eq.~(\ref{eq12}). At $t=1$ and $t=2$ the degree of the nodes is a better predictor of cascade size than the eigenvector centrality or nonbacktracking centrality (the blue line is lower than the black and red lines in panels (a) and (b)), but this situation is reversed at later times (black and red lines are lower than the blue lines in panels (c) and (d)). In all cases the tree size influence metric defined in Sec.~\ref{sec:Intro} is better than (or at least as good as) every other metric.

\begin{figure}
\centering
\epsfig{figure=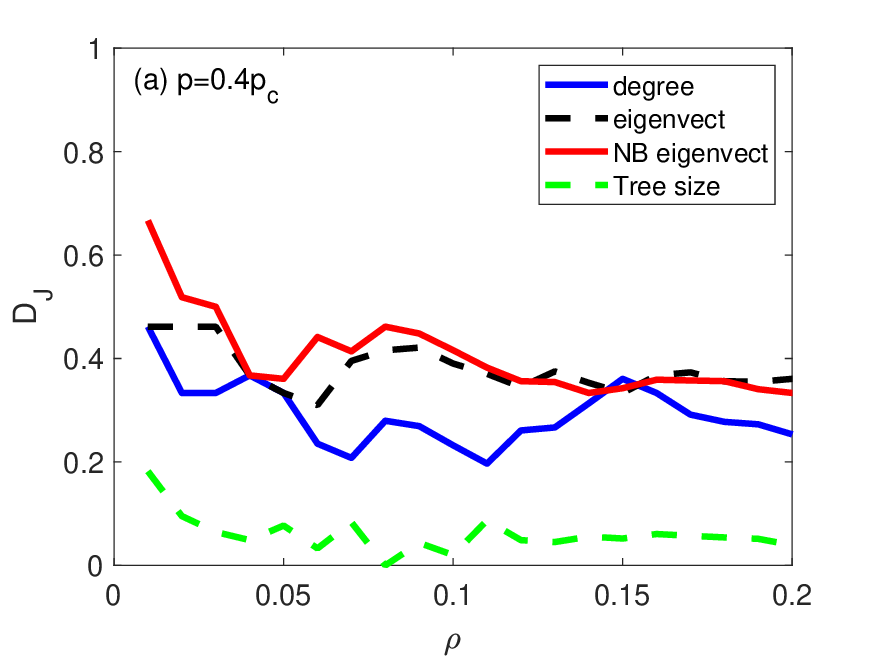,width=8.1 cm} 
\epsfig{figure=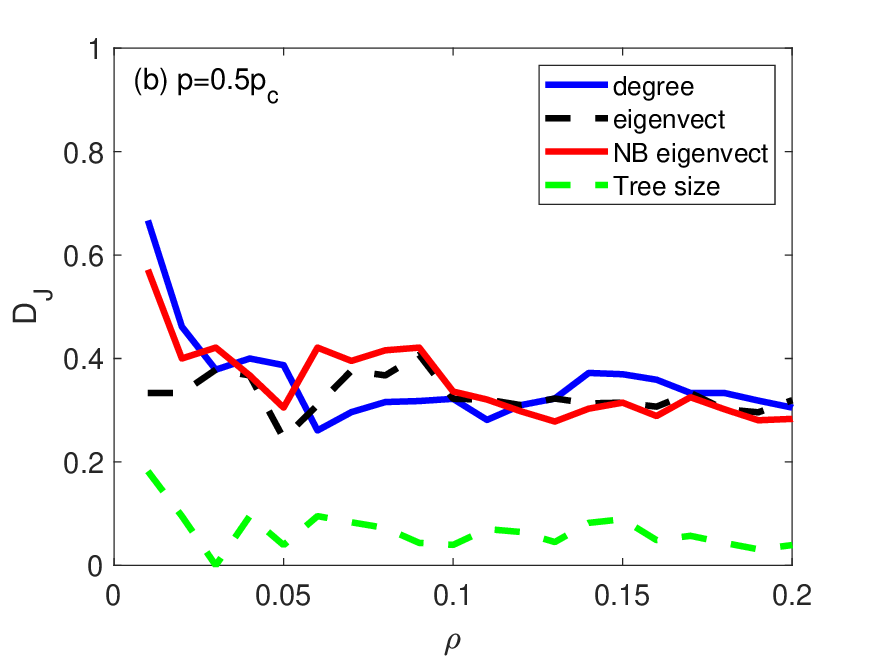,width=8.1 cm}
\epsfig{figure=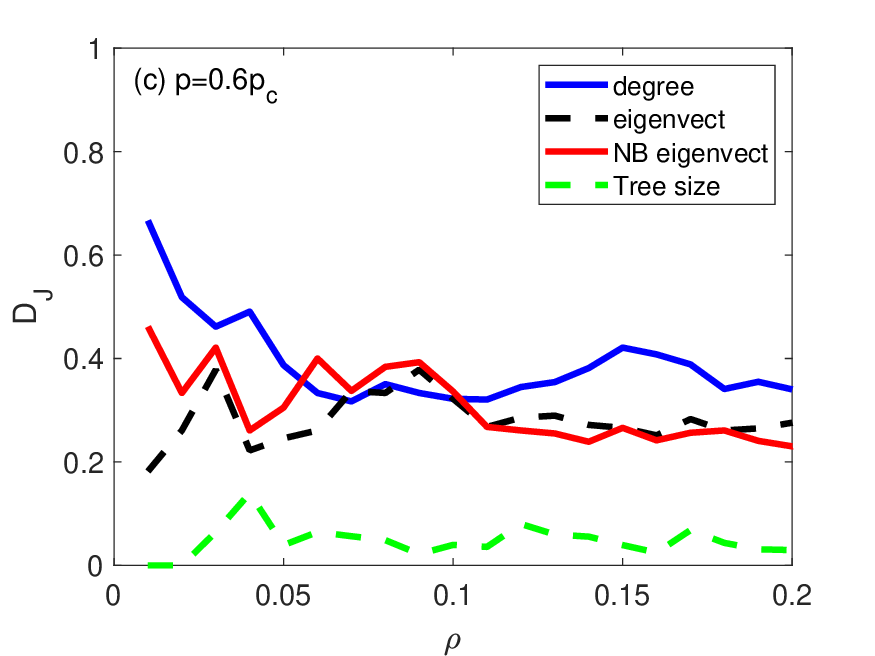,width=8.1 cm}
\epsfig{figure=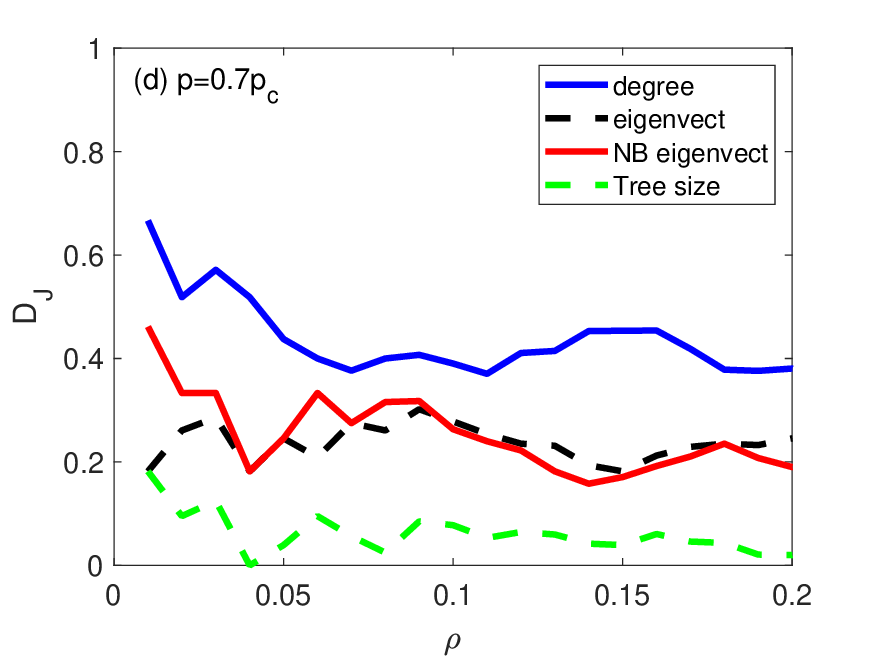,width=8.1 cm}
\epsfig{figure=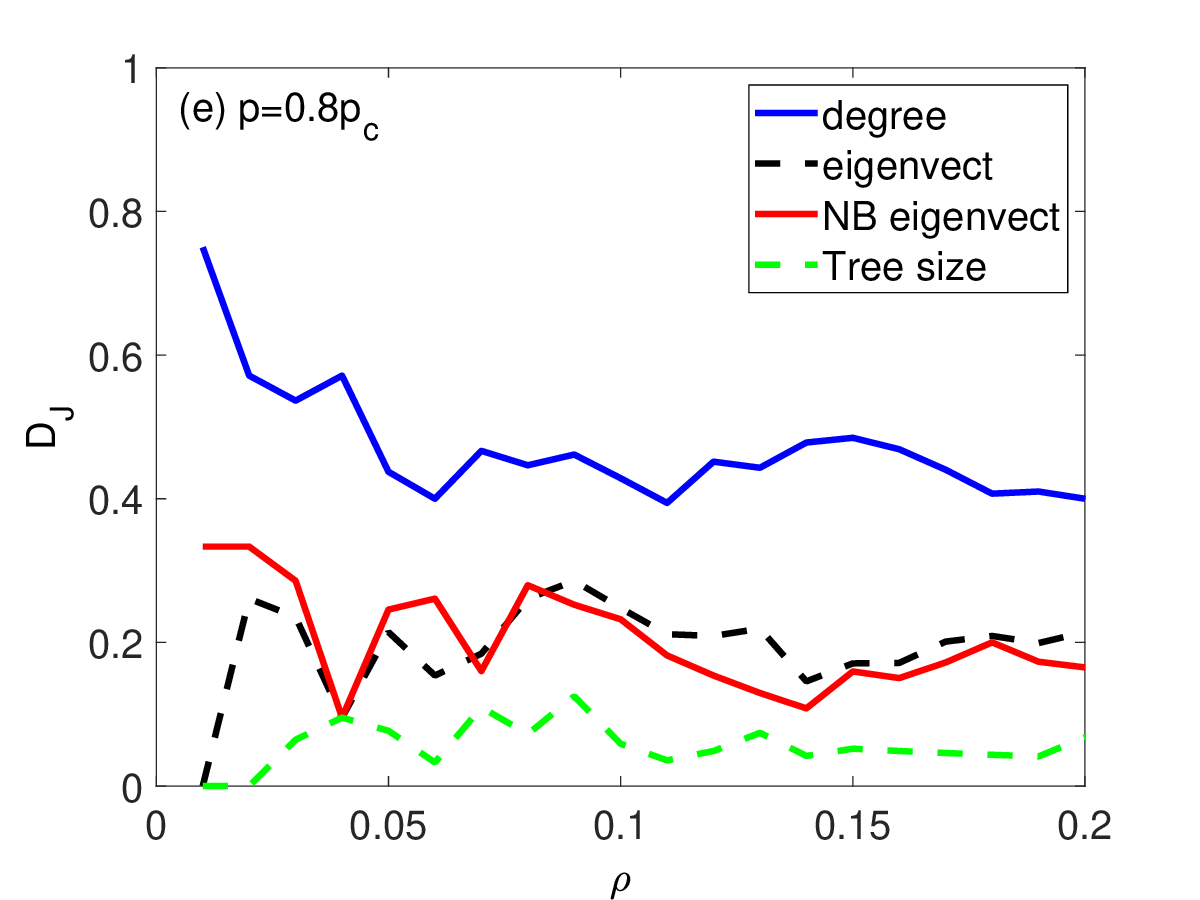,width=8.1 cm}
\epsfig{figure=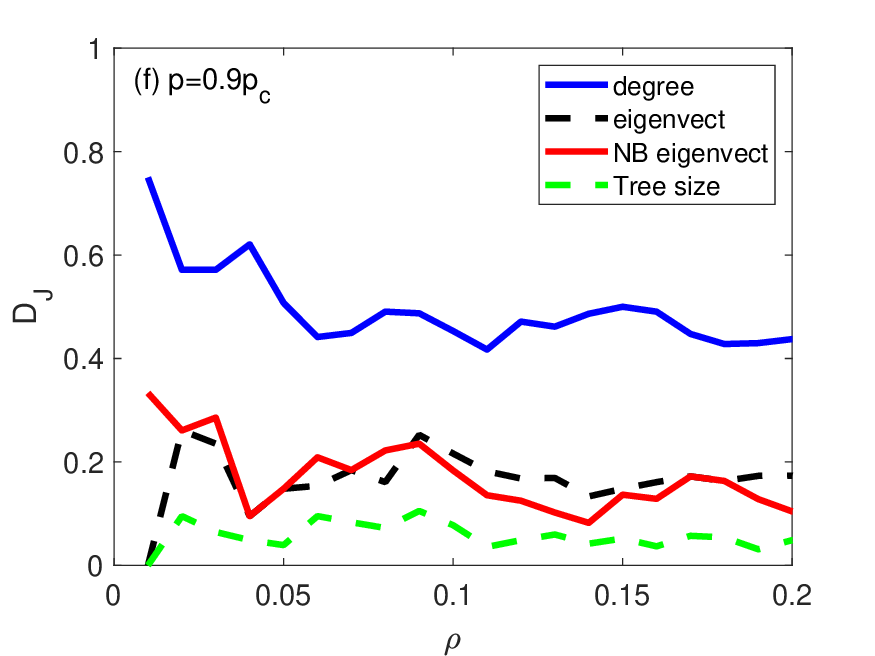,width=8.1 cm}
\caption{\textcolor{black}{Jaccard distance for the Independent Cascade Model, as in Fig.~\ref{figsubcrit}, but at time $t=10$ for a range of values of the transmission parameter 
$p$ relative to the critical value $p_c$: (a) $p=0.4p_c$, (b) $p=0.5p_c$, (c) $p=0.6p_c$, (d) $p=0.7p_c$, (e) $p=0.8p_c$, (f) $p=0.9p_c$. }}\label{figscanp}
\end{figure}

\textcolor{black}{In Figure~\ref{figscanp} we show the Jaccard distances at the same time $t=10$ for a range of $p$ values from $0.4p_c$ to $0.9p_c$. The qualitative results---and in particular the good performance of the tree size metric relative to the alternatives---is very robust to variation in the parameter $p$.}

The critical case $p=p_c$ that is the focus of \cite{Radicchi16a} is considered in Appendix~\ref{appB}, where similar results are found. In addition, the tree size influence and the nonbacktracking eigenvector centrality coincide in the $t=\infty$ limit of this critical case.

\section{Meme propagation model} \label{sec:memeprop}
We consider also the continuous-time meme propagation model of \cite{Gleeson14,Weng12} on directed networks. A meme is a distinct piece of digital information that can be copied and transmitted; here we model broadcasting social networks such as Twitter. Each node of the network represents a user, and each user holds a meme of current interest on a ``screen''. The out-neighbours of a node are its ``followers'' on Twitter: when a node sends a tweet, it is received by all its followers. Users become active at rate 1, and when a user that has a meme on her screen becomes active either  (with probability $\mu$) she creates a new meme, which appears on her screen and is tweeted to her followers, or (with probability $1-\mu$) she retweets the meme that is currently on her screen. In either case, the meme tweeted overwrites any memes that are on the screens of the node's followers. To avoid multiple retweeting of the same meme by one node, the screen of a tweeting node is set to the ``empty'' state immediately after its tweeting event (see Sec.~S5 of \cite{Gleeson14}). If a node becomes active when it has an empty screen 
 it creates a new meme and tweets this meme to its followers.

The stochastic dynamics of the model lead to some memes becoming very popular (meaning they are retweeted many times), while most memes die out quickly. The nodes who create memes that, on average, are most often retweeted are deemed the most influential nodes. Although having a large number of followers (high out-degree) is helpful to a node's influence, we see that there are also other important factors that arise from the dynamics of the information transmission process. In particular, a node's influence is affected by how likely it is that each of its followers retweets the messages it sends. In Appendix~\ref{appC} we show that the transmission rate along the directed edge $i\to j$ at a time $\tau$ after node $i$ receives the meme (i.e., becomes infected) is
\begin{equation}
\psi_{i \to j}(\tau) = (1-\mu) e^{-\left(1+\text{indeg}(j)\right)\tau}, \label{CICpsi}
\end{equation}
where $\text{indeg}(j)$ is the in-degree of node $j$, i.e., the number of users it follows on Twitter. The higher this number, the more competition there is between the incoming tweets for node $j$'s attention: recall that each incoming tweet overwrites the previous meme on the screen, so higher in-degree means that each meme has a shorter average lifetime on the node's screen, and hence less chance to be retweeted. Thus, a node with a large number of followers may not necessarily be very influential: if all those followers also follow large numbers of users, their ability to propagate messages is relatively low. A node with fewer, but more dedicated (i.e., low in-degree) followers could potentially therefore be more influential that a node with more followers.

\begin{figure}
\centering
\epsfig{figure=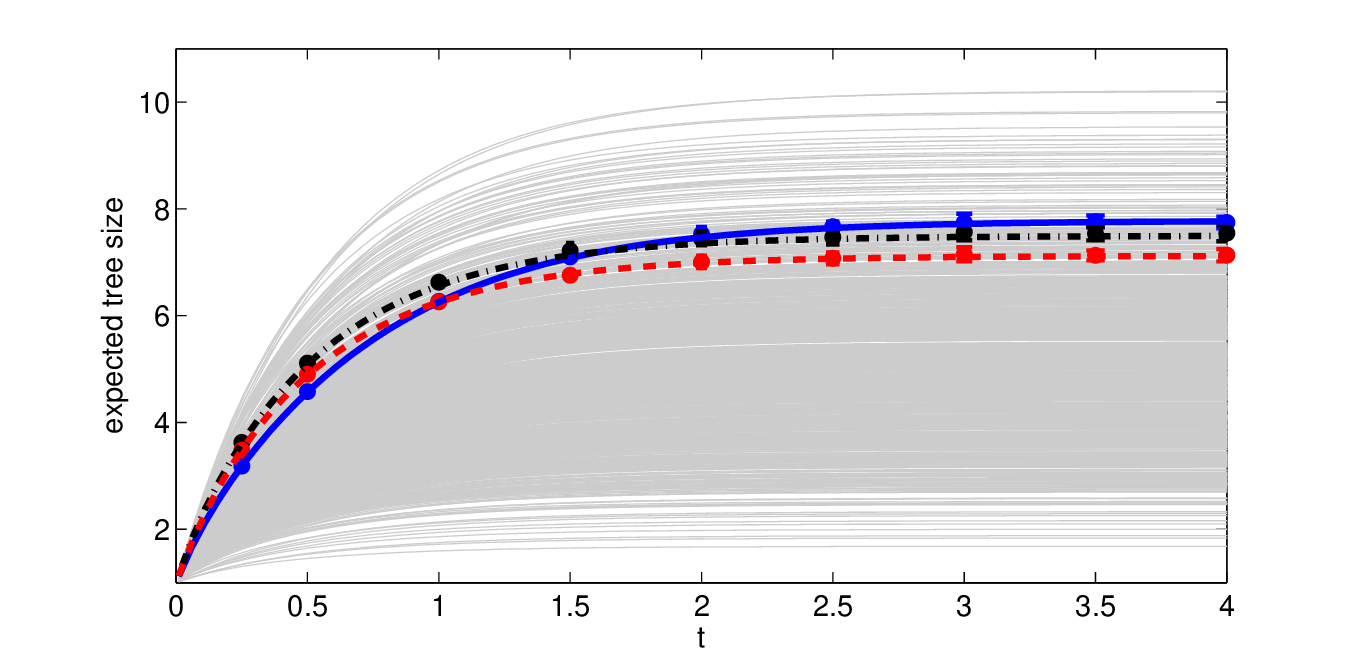,width=12.4 cm} 
\epsfig{figure=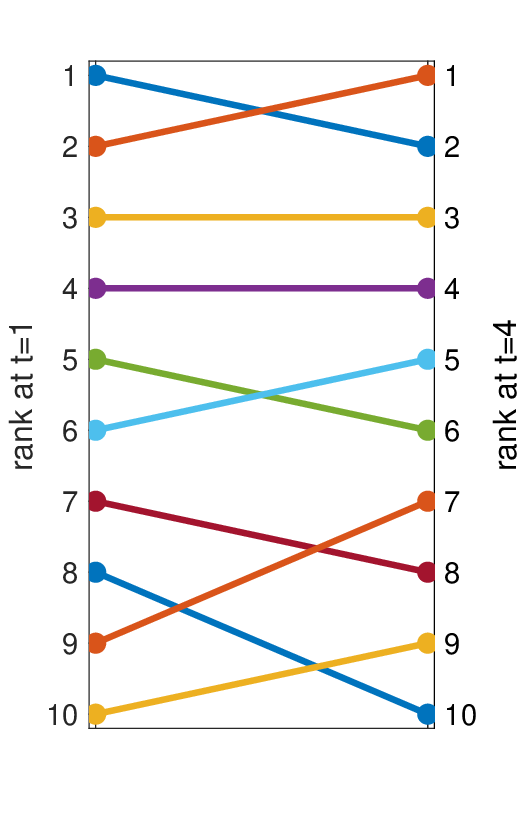,width=3.9cm}
\caption{Time-dependent tree sizes $s_i(t)$ for each of $10^3$ nodes in a single directed Erd\H{o}s-R\'{e}nyi network, using the meme propagation model described in Sec.~\ref{sec:memeprop}.  The blue, red, and black curves highlight the influences of three chosen nodes. Note that the influence of the red node and the black node are higher than that of the blue node at early times, but the blue node is more influential than the other two when cascades run to longer times. Symbols are the results of numerical simulations of the model, with error bars as described in the text. \textcolor{black}{Right: rankings of the top-10 seed nodes in terms of cascade sizes at time $t=1$ (left) and at time $t=4$ (right); each line links the rankings of a node at the two times. The node with largest cascade size at $t=1$ (rank 1 on left axis) has the second-largest cascade size at $t=4$ (rank 2 on right axis), and such crossovers are relatively common.}}\label{figtimedep}
\end{figure}
Our predictions for the tree-size influence metric applied to this model are determined by solving the linear system Eq.~(\ref{LT}) for the Laplace transform $\widehat m_{i\to j}(s)$, numerically inverting the transform (see Appendix~\ref{appD}) to obtain $m_{i\to j}(t)$, and then using Eq.~(\ref{2}) to obtain $s_i(t)$. As an example, we use a single directed Erd\H{o}s-R\'{e}nyi network with $N=10^3$ nodes, setting the innovation probaility to $\mu=0.1$. The time-dependent tree-size predictions of the theory for all nodes are shown by the grey curves in Fig.~\ref{figtimedep}: the influence of each node grows over time as the memes it seeds propagate through the network. However, note that the influence of nodes grows at different rates. We highlight the $s_i(t)$ curves of three specific nodes (see the red, blue and black curves in Fig.~\ref{figtimedep}): notice how the red and black curves both exceed the blue curve in the early stages of propagation ($t<1$), but the blue node is the most influential of the three in the long-time limit. As we saw in the previous section, the ranking of nodes' influences depends on the time the cascade has been propagating: a node who is highly influential in the first few minutes of tweeting a message may not be the most influential when the impact of longer-time propagation is included. The symbols of each colour in Fig.~\ref{figtimedep} are the average tree sizes measured in numerical simulations of the model, which match the theoretical predictions very well at all times\footnote{We calculate the average tree size by tracking the sizes of $10^4$ retweet trees seeded by each node. As a check on robustness, we also replicate the entire experiment 24 times; the error bars on the simulation results show the standard deviation of the calculated means over the 24 replicate experiments.}.

\textcolor{black}{
In the right panel of Fig.~\ref{figtimedep} we further analyze the changes in rankings by comparing the rankings of the top-10 nodes by cascade size at two times: $t=1$ (ranks on left)  and $t=4$ (ranks on right). There are multiple crossovers where nodes swap positions in the rankings between the two time snapshots (e.g. the nodes ranked 1 and 2 at time $t=1$ have swapped their ranks by $t=4$, meaning that the most influential node at $t=1$ is only the second most influential node at $t=4$); this shows that the crossover effect is quite common.
}

\begin{figure}
\centering
\epsfig{figure=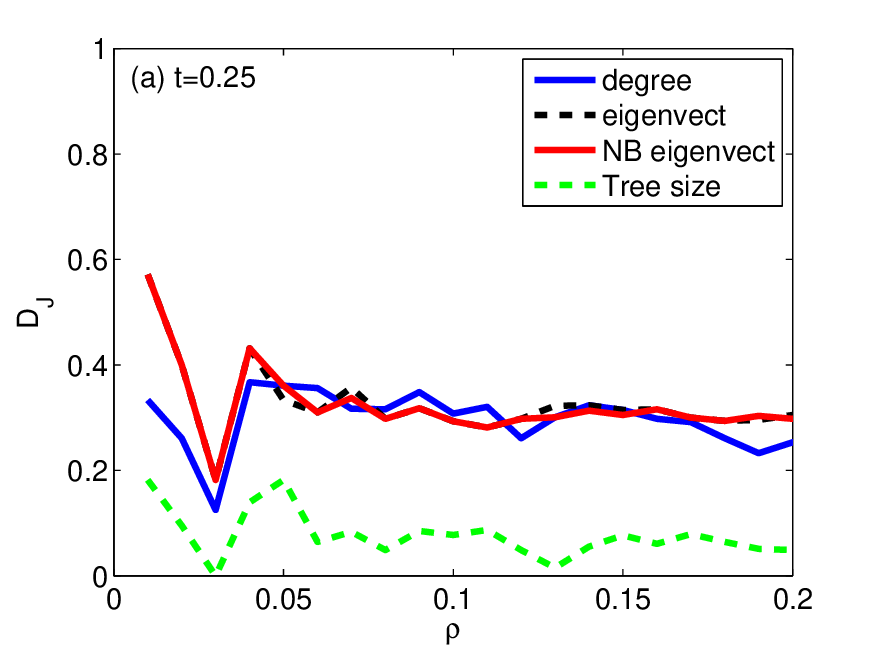,width=8.1 cm} 
\epsfig{figure=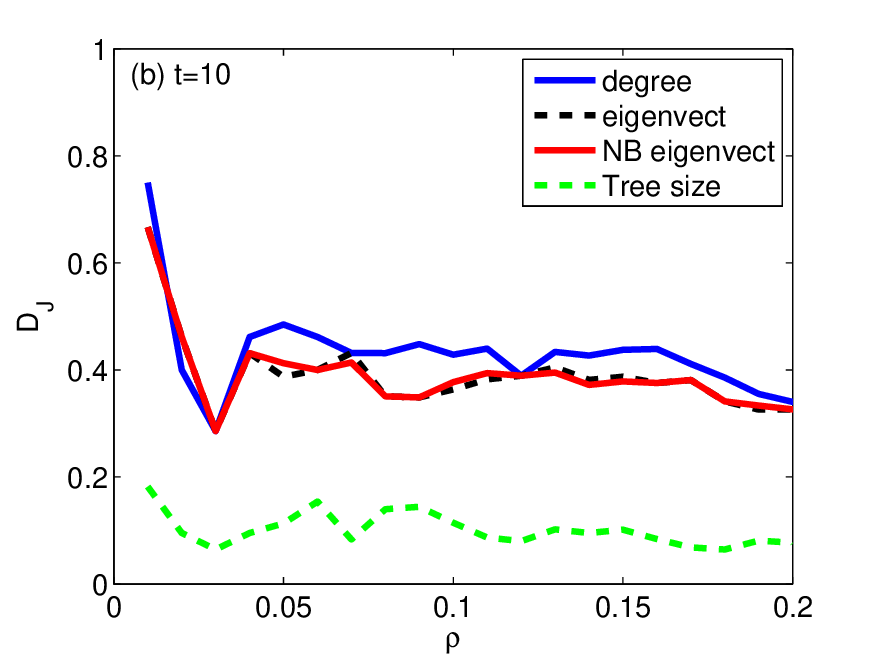,width=8.1 cm}
\caption{Jaccard distance, as in Fig.~\ref{figsubcrit}, but for the meme propagation model used in Fig.~\ref{figtimedep}.}\label{figCICER}
\end{figure}
In Figure~\ref{figCICER} we compare the accuracy of node influence rankings given by the (out-) degree centrality, eigenvector centrality, and non-backtracking centrality with our tree-size metric, using the Jaccard distance $D_J$ in the same way as in Fig.~\ref{figsubcrit}. The excellent prediction of the tree size metric is reflected in the low Jaccard distances between the theoretical and actual node rankings; the quality of the prediction is consistent over all timescales of the cascades. The relatively poor performance of the structure-based metrics is unsurprising in this case, because the probability $p_{i\to j}$ that the infection of node $i$ leads to the infection of node $j$ is given by Eqs.~(\ref{vpeqn}) and (\ref{CICpsi}) as
\begin{equation}
p_{i \to j} = \frac{1-\mu}{1+\text{indeg(j)}}.
\end{equation}
Unlike the epidemic model of Sec.~\ref{sec:ICM}, where the value of $p_{i\to j}$ was the same for all edges, here it depends on the number of nodes followed by $j$ (as a reflection of the limited attention of Twitter users and consistent with empirical results \cite{Lerman16,Weng12}). This effect of the transmission dynamics is ignored in structure-based metrics that consider only connectivity or paths, but it plays a crucial role in determining the relative influence of the nodes for information transmission.

\section{Real-world networks} \label{sec:realworld}
The examples in Secs.~\ref{sec:ICM} and \ref{sec:memeprop} used synthetic networks. In this section we examine the performance of the tree-size influence metric on real-world networks, which are not locally tree-like.

\begin{figure}
\centering
\epsfig{figure=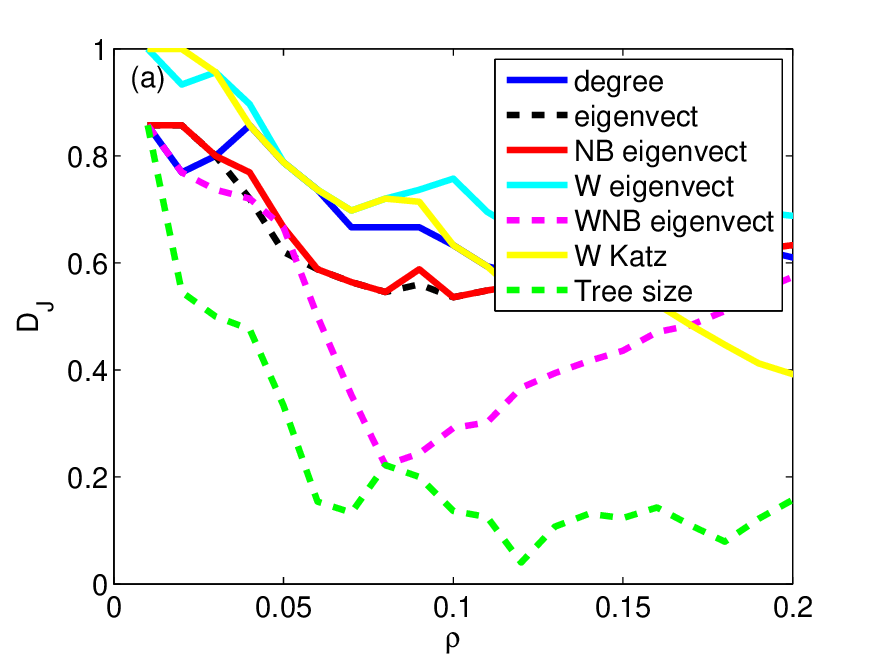,width=8.1 cm} 
\epsfig{figure=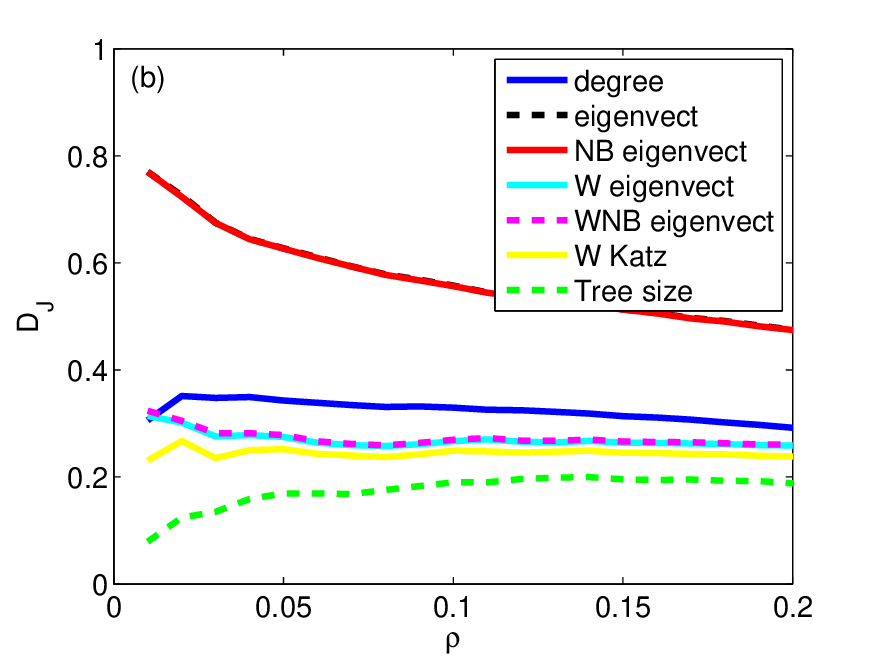,width=8.1 cm}
\caption{Jaccard distance for various influence metrics on real-world networks: (a) Independent Cascade Model with edge-dependent transmission probabilities $p_{i j}$, as determined from the face-to-face contact data of $N=410$ people in \cite{Sociopatterns1,Sociopatterns2}; (b) Meme propagation model on the Twitter network of the Spanish 15M protest movement \cite{gonzalez11,borge11}.}\label{figrealworld}
\end{figure}

The first example uses the face-to-face contact data of $N=410$ people who visited the Science Gallery in Dublin on one day\footnote{The Science Gallery data was collected on July 15, 2009.} \cite{Sociopatterns1,Sociopatterns2}. Using the data on the contact times, we construct a weighted undirected network, where the weight on the edge between nodes $i$ and $j$ is proportional to the amount of time they spent in contact with each other. We use these weights to generate infection probabilities $p_{i j}=p_{j i}$ for each pair \textcolor{black}{$i$ and $j$} of nodes and then simulate the dynamics using the Independent Cascade Model, but using the edge-dependent probabilities $p_{i j}$ for transmission along the $i$--$j$ edge instead of the single parameter $p$ of Sec.~\ref{sec:ICM}. Figure~\ref{figrealworld}(a) shows the Jaccard distances for various influence metrics, compared with the average outbreak sizes from numerical simulations ($10^4$ cascades seeded by each node; sizes are evaluated when all spreading has ceased, i.e., $t=\infty$). In addition to the four structural metrics already described in Sec.~\ref{sec:ICM}, we also consider here metrics that incorporate the weights $p_{i j}$. The weighted eigenvector centrality measure (``W eigenvect'') ranks nodes according to the corresponding values in the leading eigenvector of the weighted adjacency matrix of the network (i.e., the matrix with entry $p_{i j}$ if nodes $i$ and $j$ meet, and zero otherwise). Similarly, the weighted non-backtracking centrality (``WNB eigenvect'') uses the leading eigenvector of the non-backtracking matrix of the weighted adjacency matrix. The weighted Katz centrality (``W Katz'') calculates the Katz centrality using the weighted adjacency matrix as \cite{Newmanbook} $\mathbf{x}=\left(\mathbf{I}_N - \mathbf{A}\right)^{-1}\mathbf{1}$, where $\mathbf{A}$ has entries $A_{i j}=p_{i j}$ and $\mathbf{x}$ is the vector of node centralities, cf.~Eq.~(\ref{hvect}).

The additional information provided by the weight $p_{i j}$ of each edge enables some of the metrics in Fig.~\ref{figrealworld}(a) (notably, the weighted non-backtracking centrality) to achieve lower Jaccard distances than the purely structure-based metrics. Nevertheless, the tree-size influence metric performs best of all the metrics. This demonstrates that our theoretical approach (which assumes an infinite-sized and tree-like network) can yield useful influence measures even for rather small real-world networks ($N=410$ here).

\textcolor{black}{It is noticeable that the tree-size Jaccard distance in Fig.~\ref{figrealworld}(a) is unusually high for the lowest values of $\rho$. Upon investigation, we find that these high values are mainly caused by a misranking of three specific nodes. These three nodes form a triangle, and each node in the triangle has a high infection probability for the other two. The tree-like assumption (that subtrees in a cascade are independent of each other) that we used to derive Eq.~(\ref{1}) is violated in this case, as the subtrees of the three nodes in the triangle strongly affect each other. The result is that the tree-size metric gives an inaccurately high ranking to the three nodes in the triangle, which affects the $D_J$ values when only the top influencers are considered (i.e., at the lowest values of $\rho$). Dealing accurately with high-strength triangles remains a challenge for future work in this direction, but we speculate that methods used to extend message-passing approaches to networks with loops \cite{cantwell2019message} may prove instructive.}

Another example of dynamics on a real-world network structure is shown in  Fig.~\ref{figrealworld}(b), which compares influence metrics for the meme propagation model described in Sec.~\ref{sec:memeprop} (with $\mu=0.3$) running on the real Twitter network of the Spanish 15M movement \cite{borge11,gonzalez11}. This network contains $N=87,569$ nodes, with mean degree $z=69$, and $44\%$ of the directed links are reciprocated (i.e., node $j$ follows node $i$ on Twitter and node $i$ also follows node $j$). Despite the non-treelike character of this network, the tree-size influence metric again outperforms all the other metrics in identifying the nodes that seed the largest cascades.

\section{Conclusions} \label{sec:conclusions}
We have introduced a measure for the influence of a node that is based on the expected size of cascades seeded from that node. The dynamics of the spreading process are incorporated through the time-dependent rate  $\psi_{i \to j}(\tau)$ of transmission from node $i$ to node $j$, where $\tau$ is the time elapsed since the infection of node $i$. We have shown that the influence of nodes may be calculated in a numerically efficient manner through the solution of sparse linear systems of equations, thus enabling the scaling of the method to large networks.

Using large-scale numerical simulations on both synthetic and real-world networks, we demonstrated that the tree-size influence metric gives a better ranking of nodes that other standard approaches, and that it reduces to the non-backtracking centrality of \cite{Radicchi16a} in the case of disease-spread dynamics with infection parameters that are precisely at the critical point between subcritical and supercritical regimes. We investigated the time-dependence of the influence metric using the discrete-time Independent Cascade Model and a continuous-time model for meme propagation. Our metric accurately captures the change in the relative influence of nodes over the duration of a cascade: this may be important on Twitter, for example, where the top influencers may differ depending on whether the impact required has a long or a short time horizon (e.g., for viral marketing or for the spreading of emergency response information, respectively). \textcolor{black}{We believe that this framework will open the way to further interesting questions; some examples include the impact of network structure (e.g. scale-free networks) on the crossover effect seen in Fig.~\ref{figtimedep} or the possible connection to methods for supercritical dynamics such as in \cite{min2018identifying}. }

\section*{Acknowledgements}
Helpful discussions with Tim Rogers, David O'Sullivan, Ioannis Dassios, Alan Hegarty and Peter Fennell are gratefully acknowledged. We acknowledge the SFI/HEA Irish Centre for High-End Computing (ICHEC) for the provision of computational facilities. This publication has emanated from research conducted with the financial support of Science Foundation Ireland under Grant numbers 16/IA/4470 (A.C., A.F.~and J.P.G.) and 12/RC/2289 P2 (J.P.G.). For the purpose of Open Access, the authors have applied a CC BY public copyright licence to any Author Accepted Manuscript version arising from this submission.

\appendix
\renewcommand{\thefigure}{S\arabic{figure}}
\setcounter{figure}{0}
\section{Appendix A: Efficient solution methods for linear systems}\label{appA}
Equation (\ref{matrixeq}) is a system of linear equations for the $E$ components of the vector $\mathbf{m}$. The $E\times E$ matrix $\mathbf{I}_E-\mathbf{P}\mathbf{B}$ is sparse---it contains at most $(z+1)E$ nonzero entries, where $z$ is the mean degree of the directed network\footnote{The identity matrix $\mathbf{I}_E$ has $E$ non-zero entries, while the nonbacktracking matrix $\mathbf{B}$ has at most $z E$ non-zero values, corresponding to the case where no reciprocal edges exist. Since matrix $\mathbf{P}$ is diagonal, the product $\mathbf{P} \mathbf{B}$ also has at most $z E$ non-zero values.}---and so efficient numerical solution methods can be used even for large matrices.

A further improvement in the speed of the numerical solution can be obtained by decomposing the nonbacktracking matrix as follows:
\begin{equation}
\mathbf{B}=\mathbf{T}\mathbf{F} - \mathbf{R}.\label{decomp1}
\end{equation}
Here $\mathbf{T}$ is the $E\times N$ matrix with entry $T_{\alpha, i}=1$ if edge $\alpha$ is an in-edge of node $i$, $\mathbf{F}$ is the $N\times E$ matrix with entry $F_{i,\alpha}=1$ if edge $\alpha$ is an out-edge of node $i$, and $\mathbf{R}$ is the $E\times E$ matrix with entry $R_{\alpha,\beta}=1$ if edges $\alpha$ and $\beta$ are reciprocal (i.e., if  $\alpha$ is the directed edge $i\to j$ and $\beta$ is the directed edge $j\to i$); otherwise all entries of the three matrices are zero\footnote{Note that the usual $N\times E$ incidence matrix of the network can be written as $\mathbf{F}-\mathbf{T^\top}$.}. Note that the $\mathbf{T}$ matrix has one nonzero entry in each row, while the $\mathbf{F}$ matrix has one nonzero entry in each column, so each matrix has $E$ nonzero entries in total. The maximum number of nonzero entries in the reciprocal matrix $\mathbf{R}$ is $E$; this occurs in the case where the original matrix is undirected, so that every directed edge has a reciprocal edge.

We define the $N$-vector $\mathbf{q}$ by
\begin{equation}
\mathbf{q}=\mathbf{F}\mathbf{m}.\label{decomp2}
\end{equation}
Then, using Eq.~(\ref{decomp1}), Eq.~(\ref{matrixeq}) can be written as
\begin{equation}
\left(\mathbf{I}_E + \mathbf{P}\mathbf{R}\right)\mathbf{m} - \mathbf{P}\mathbf{T}\mathbf{q} = \mathbf{p}. \label{decomp3}
\end{equation}
Equations (\ref{decomp2}) and (\ref{decomp3}) can be combined to give a $(N+E)\times (N+E)$ system of linear equations as follows:
\renewcommand{\arraystretch}{1.5}
\begin{equation}
\left( \begin{array}{cc}
        \mathbf{I}_N & -\mathbf{F} \\
        -\mathbf{P T} &  \mathbf{I}_E+\mathbf{P R}
        \end{array} \right)
\left( \begin{array}{c}
        \mathbf{q}\\
        \mathbf{m}
        \end{array} \right)
=
\left( \begin{array}{c}
        \mathbf{0}\\
        \mathbf{p}
        \end{array} \right).\label{decomp4}
\end{equation}
Although the dimensions of this system are larger than those of Eq.~(\ref{matrixeq}), which is $E\times E$, the matrix on the left hand side of Eq.~(\ref{decomp4}) is (for most networks of interest), significantly more sparse than the matrix of Eq.~(\ref{matrixeq}), which results in faster solution of the linear system. Specifically, the number of nonzero entries in the matrix of Eq.~(\ref{decomp4}) is less than or equal to $4E+N$ (the subblock $\mathbf{I}_N$ contributes $N$ nonzero entries, while each of $\mathbf{F}$ and $\mathbf{P T}$ contributes $E$ (recall that $\mathbf{P}$ is diagonal), and the maximum number of nonzero entries from the $\mathbf{I}_E+\mathbf{P R}$ subblock is $2E$). Writing $N=E/z$ and comparing with the $(z+1)E$ nonzero entries calculated above for the matrix of Eq.~(\ref{matrixeq}), we conclude that the block matrix in Eq.~(\ref{decomp4}) has fewer nonzero entries than that of Eq.~(\ref{matrixeq}) if the mean number $z$ of directed edges per node exceeds $3.3$. This is the case for most networks of interest; indeed, online social networks often have values of $z$ on the order of hundreds, which makes the use of the decomposition method highly advantageous.

\section{Appendix \textcolor{black}{B}: Numerical inversion of Laplace transforms} \label{appD}
The linear system of Eq.~(\ref{LT}) can be written in a form similar to Eq.~(\ref{matrixeq}) as
\begin{equation}
\left(\mathbf{I}_E - \mathbf{\widehat P}(s)\mathbf{B}\right) \mathbf{\widehat m}(s) = \mathbf{\widehat p}(s), \label{LTapp1}
\end{equation}
where $\mathbf{\widehat p}(s)$ is the vector of $\frac{1}{s}\widehat{\psi}_{i \to j}(s)$ values and $\mathbf{\widehat P}(s)$ is the $E\times E$ matrix with the values $\widehat{\psi}_{i\to j}(s)$ on the diagonal (and zeros elsewhere). For any given value of the Laplace transform variable $s$, the system (\ref{LTapp1}) can be solved efficiently using the methods described in Appendix~\ref{appA}. Then we calculate the inverse Laplace transform using the Talbot inversion formula \cite{Abate06}:
\begin{equation}
\mathbf{m}(t) = \frac{2}{5 t}\sum_{k=0}^{M-1}\text{Re}\left[ \gamma_k\, \mathbf{\widehat m}\left(\frac{\delta_k}{t}\right)\right], \label{LTapp2}
\end{equation}
where the complex-valued parameters are given by \cite{Abate04}
 \begin{align}
 \delta_0&= \frac{2M}{5}, \quad \delta_k=\frac{2k\pi}{5}(\cot(k\pi/M)+i)\quad\text{ for }0 <k< M,\nonumber\\
 \gamma_0&=\frac{1}{2}e^{\delta_0},\quad
 \gamma_k=[1+i(k\pi/M)(1+[\cot(k\pi/M)^2])-i\cot(k\pi/M)]e^{\delta_k}\quad\text{ for }0< k<M,
 \end{align}
with $i=\sqrt{-1}$.

The procedure to determine $\mathbf{m}(t)$ at a desired time value $t$ is to solve the linear system of Eq.~(\ref{LTapp1}) for $\mathbf{\widehat m}(s)$ at each of the $M$ values of $s$ given by $\delta_k/t$ for $k=0$ to $M-1$. Substituting these vectors into the inversion formula of Eq.~(\ref{LTapp2}) gives the value of $\mathbf{m}(t)$. The Talbot algorithm gives very high accuracy even for low values of $M$; we use $M=10$ in our calculations for Figure~\ref{figtimedep}.

\section{Appendix \textcolor{black}{C}: Independent Cascade Model at criticality} \label{appB}
\begin{figure}
\centering
\epsfig{figure=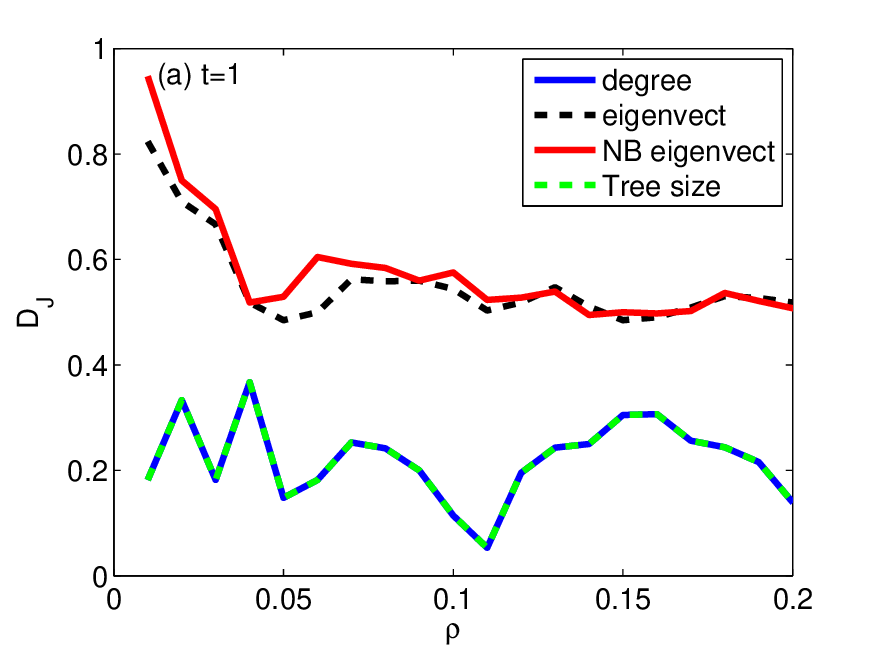,width=8.1 cm} 
\epsfig{figure=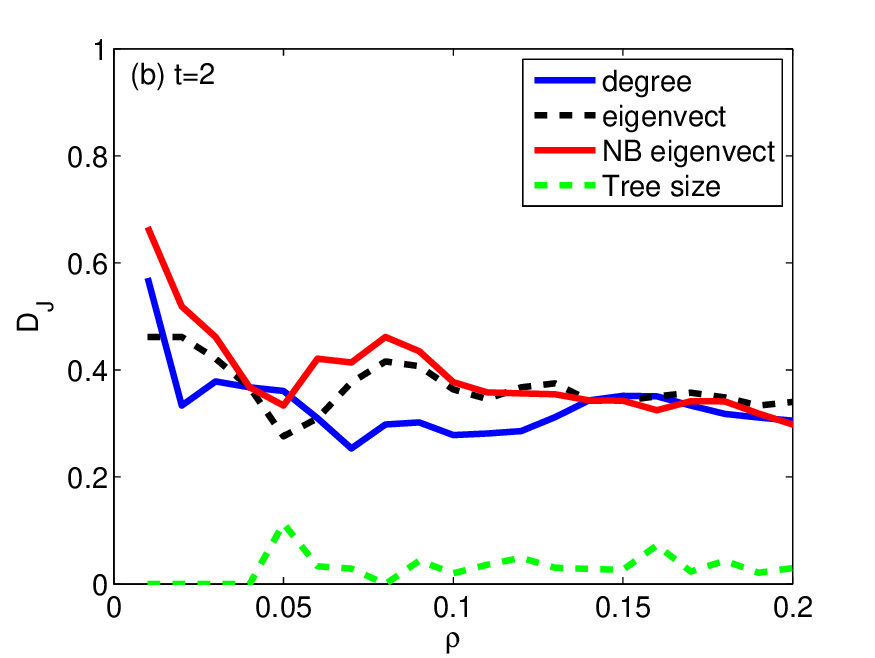,width=8.1 cm}
\epsfig{figure=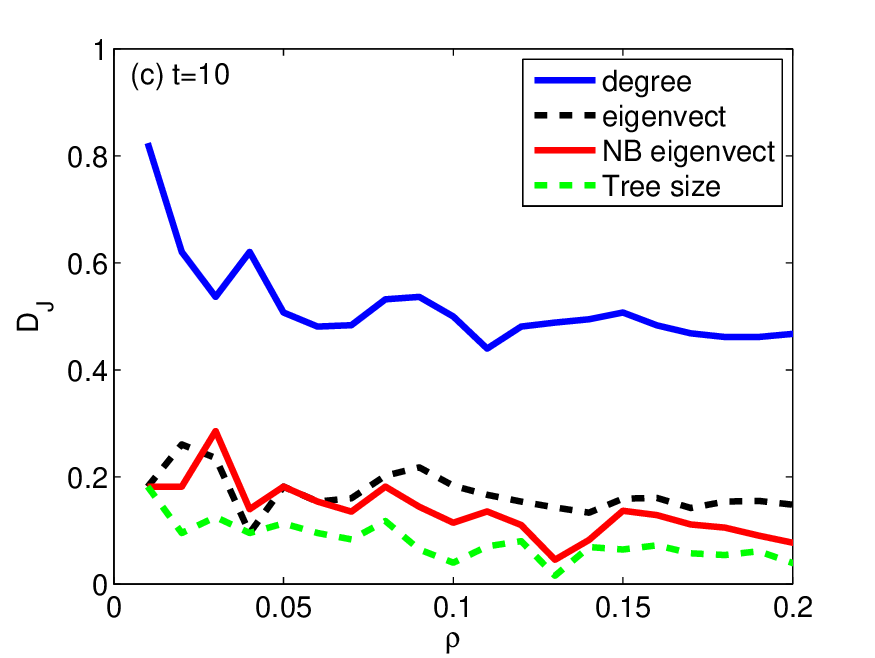,width=8.1 cm}
\epsfig{figure=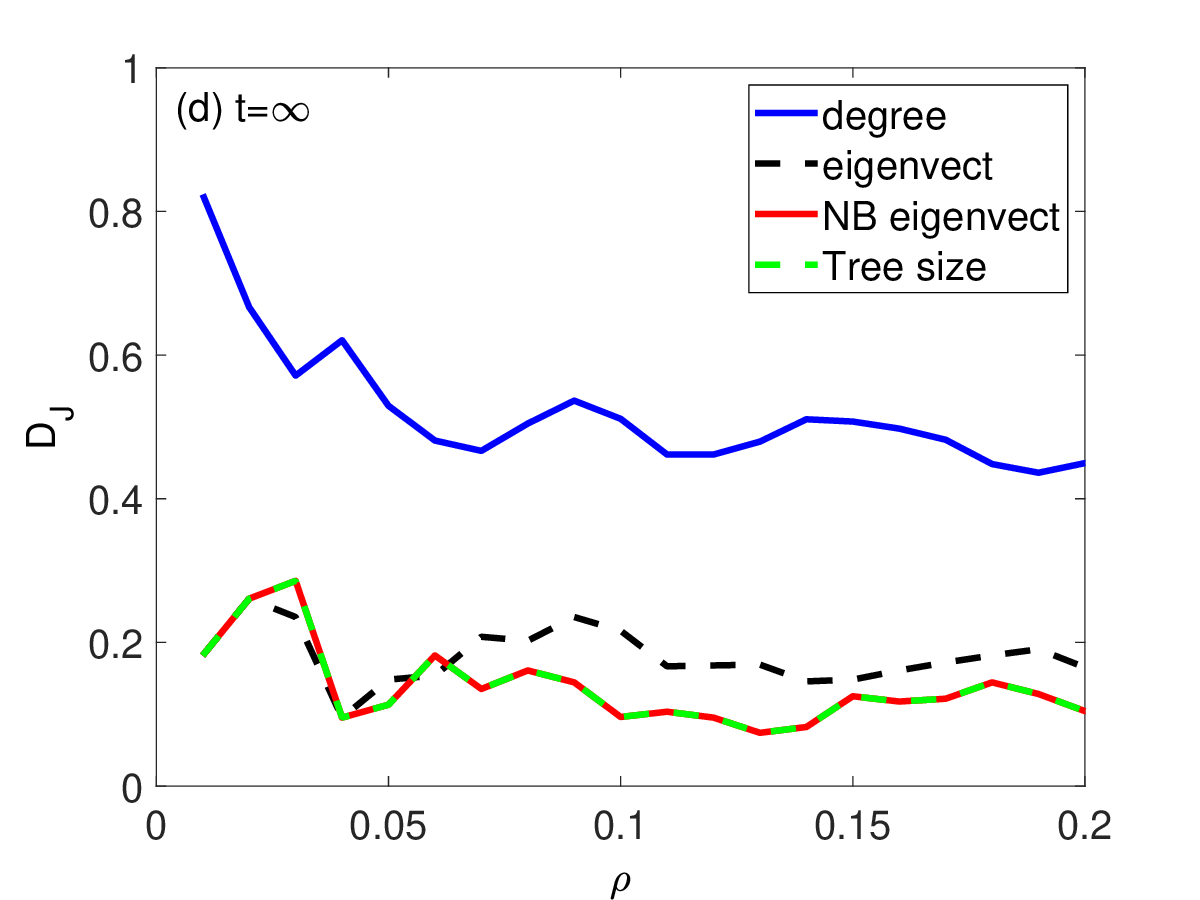,width=8.1 cm}
\caption{As Figure~\ref{figsubcrit} but for the critical case $p=p_c$. \textcolor{black}{Note that the tree size influence coincides with the nonbacktracking eigenvector centrality in the case $t=\infty$ (panel (d)), which explains why the nonbacktracking centrality is found to be the most accurate influence metric for SIR dynamics near criticality in \cite{Radicchi16a}}. }\label{figcrit}
\end{figure}

Figure~\ref{figcrit} shows results for the Independent Cascade Model on the same network as used for Fig.~\ref{figsubcrit} of the main text, but with the transmission probability equal to its critical value: $p=p_c$. The results display a similar pattern to those in Fig.~\ref{figsubcrit}: again we see that the tree size influence is equivalent to degree centrality at $t=1$, that the degree centrality loses accuracy relative to the eigenvector and nonbacktracking centralities as $t$ increases, and that the tree size influence metric is the most accurate (or joint best) in all cases. In this critical case we also note that the tree size influence and the nonbacktracking eigenvector centrality coincide in the $t=\infty$ limit, so that the results of \cite{Radicchi16a} are a special case of our time-dependent influence metric.

\section{Appendix \textcolor{black}{D}: Derivation of Eq.~(\ref{CICpsi})} \label{appC}
Consider the meme being tweeted by node $i$ at a time that we call $\tau=0$\footnote{In the meme propagation model, ``infection of node $i$ at time $\tau$'' means ``node $i$ tweets the meme of interest at time $\tau$''.}. A follower $j$ of node $i$ thus receives the meme onto his screen at $\tau=0$, and we want to determine the time-dependent probability that node $j$ retweets the meme thereafter. Note that a  retweet, if it occurs, must happen before any other in-neighbour of node $j$ tweets, since a tweeting in-neighbour causes the meme on node $j$'s screen to be overwritten, which makes it impossible for $j$ to retweet that meme. In this Markovian model, the rate at which node $j$ retweets (given that he has a non-empty screen) is $1-\mu$, and since this is a Markovian model, the waiting time $t$ until node $j$'s next retweet event is a random variable with the exponential distribution
of mean $1/(1-\mu)$, given by
\begin{equation}
P_\text{retweet}(t)=(1-\mu)e^{-(1-\mu)t}.
\end{equation}

The meme on node $j$'s screen is overwritten at rate $\text{indeg}(j)+\mu$, since each in-neighbour tweets at rate 1, and in addition node $j$ innovates at rate $\mu$; both cases lead to the current meme being overwritten. Hence, the distribution of the overwriting times is
\begin{equation}
P_\text{overwrite}(t)=\left(\text{indeg}(j)+\mu\right)e^{-\left(\text{indeg}(j)+\mu\right)t}.
\end{equation}
The probability that  node $j$ will retweet the meme on its screen before it is overwritten is therefore given by \cite{Karrer10}
\begin{align}
\psi_{i\to j}(\tau)\,d\tau &= P_\text{retweet}(\tau)\,d\tau \int_\tau^\infty P_\text{overwrite}(t) dt\\
& = (1-\mu) e^{-(1-\mu)\tau} e^{-\left(\text{indeg}(j)+\mu\right)\tau}\,d\tau,
\end{align}
which yields Eq.~(\ref{CICpsi}).

\bibliographystyle{unsrt}
\bibliography{influence_refs}

\end{document}